\documentstyle[prd,aps,preprint]{revtex}

\newcommand{\bdi}{\begin{displaymath}}
\newcommand{\edi}{\end{displaymath}}
\newcommand{\bfi}{\begin{figure}}
\newcommand{\efi}{\end{figure}}

\newcommand{\beq}{\begin{equation}}
\newcommand{\eeq}{\end{equation}}

\newcommand{\gam}{\gamma_{\mu}}

\newcommand{\gaf}{\gamma_{5}}
\newcommand{\beqa}{\begin{eqnarray}}
\newcommand{\eeqa}{\end{eqnarray}}

\newcommand{\ra}{\rightarrow}

\newcommand{\wt}{\widetilde}

\newcommand{\dsla}{\partial\hspace{-6pt} /  }  
\newcommand{\Asla}{A\hspace{-6.5pt}  /  } 
\begin{document}

\def\footnoterule{\kern-3pt \hrule width\hsize \kern3pt}
\tighten
\title{INVESTIGATION OF ANOMALOUS AXIAL QED\thanks{This 
work is supported by a Schr\"odinger Stipendium of the Austrian FWF and
in part by funds provided by the U.S.
Department of Energy (D.O.E.) under cooperative 
research agreement \#DF-FC02-94ER40818.}}
\author{Christoph Adam\footnote{Email address: {\tt adam@ctp.mit.edu, 
adam@pap.univie.ac.at}}}

\address{Center for Theoretical Physics \\
Laboratory for Nuclear Science \\
and Department of Physics \\
Massachusetts Institute of Technology \\
Cambridge, Massachusetts 02139 \\
and \\
Inst. f. theoret. Physik d. Uni Wien \\
Boltzmanngasse 5, 1090 Wien, Austria \\
{~}}

\date{MIT-CTP-2617,~  {~~~~~} March 1997}
\maketitle
\thispagestyle{empty}

\begin{abstract}
Although axial QED suffers from a gauge anomaly, gauge invariance
may be maintained by the addition of a nonlocal counterterm. Such
nonlocal conterterms, however, are expected to ruin unitarity
of the theory. We explicitly investigate some relevant Feynman diagrams
and show that, indeed, unitarity is violated, contrary to recent
claims.

\medskip

\end{abstract}
\vspace*{\fill}
%\begin{center}
%Submitted to: {\it Research Journal}
%\end{center}


\input psbox.tex
\let\fillinggrid=\relax
\section{Introduction}

We want to investigate axial QED in four dimensions, 
i.e. the theory of one massless 
fermion coupled to a gauge field $A_\mu$ via the axial current
$J^5_\mu =\bar\Psi \gam\gaf \Psi$. The Lagrangian reads (we use the 
conventions $\gaf = i\gamma^0 \gamma^1 \gamma^2 \gamma^3$ and
$\epsilon_{0123} =1$)
\beq
L=\bar\Psi (i\dsla +e\Asla \gaf )\Psi -\frac{1}{4}F_{\mu\nu}F^{\mu\nu}
\eeq
The generally accepted point of view is that this theory may not be
quantized consistently because of the axial anomaly \cite{Ad1} -- \cite{Ja1}
that spoils gauge
invariance \cite{GJ1}. 
It is, however, wellknown that the effective gauge field action
\beq
e^{iS_{\rm eff}[A_\mu]}=\int D\bar\Psi D\Psi e^{iS[\Psi ,\bar\Psi ,A_\mu]}
\eeq
may be made gauge invariant by the addition of a nonlocal counterterm
(see e.g. \cite{Ber,Kieu})
\beq
S_{\rm ct} =\frac{e^3}{48\pi^2}\epsilon^{\mu\nu\alpha\beta}\int d^4 xd^4 y
\partial^\lambda_x A_\lambda (x) \Box^{-1} (x-y)F_{\mu\nu}(y)
F_{\alpha\beta}(y)
\eeq
Via the axial anomaly
\beq
\partial^\mu J^5_\mu =\frac{e^2}{48 \pi^2}\epsilon^{\mu\nu\alpha\beta}
F_{\mu\nu}F_{\alpha\beta}
\eeq
this counterterm changes the interaction to
\beq
S_{\rm I} =e\int d^4 x A^\mu (x)(J^5_\mu (x) -\frac{\partial^\mu
\partial^\nu}{\Box}J^5_\nu (x))
\eeq
which makes the gauge invariance obvious.

So one could ask if by the inclusion of this counterterm into the original
action (1) a reasonable quantum theory may be obtained. Usually such nonlocal
terms are rejected because the propagator $\Box^{-1}(x-y)$ (the Feynman
propagater of a massless boson) is expected to produce additional 
contributions to imaginary parts of e.g. scattering amplitudes, 
thereby spoiling unitarity. 
But in a recent paper \cite{Fed} (see also \cite{Kieu}) it is argued that
the net contribution of this counterterm to physical processes  vanishes,
and therefore the inclusion of the counterterm leads to an acceptable
quantum theory. (Actually, in \cite{Fed} a more general model, including
both vector and axial vector gauge coupling, was investigated. It is,
however, the axial coupling that suffers from a gauge anomaly.)

In this paper we explicitly investigate some Feynman diagrams and show that
there exist physical processes where the counterterm does contribute,
and therefore unitarity is spoiled, at least perturbatively.
Our conventions are those of \cite{Ber}.

\section{Computation of the counterterm}

We study the model
\beq
L=\bar\Psi (i\dsla +e\Asla \gaf )\Psi -\frac{1}{4}F_{\mu\nu}F^{\mu\nu}
+\frac{e^3}{48\pi^2}\epsilon^{\mu\nu\alpha\beta} \int dy 
\partial^\lambda_x A_\lambda (x)\Box^{-1} (x-y)F_{\mu\nu}(y)F_{\alpha\beta}(y)
\eeq
In the theory without counterterm the anomaly stems from the lowest order
contribution to the three-point function
\beq
T_{\mu\nu\lambda}(k_1 ,k_2 ,k_3):={\rm FT}(\langle J^5_\mu (x)J^5_\nu (y)
J^5_\lambda (0)\rangle )
\eeq
\bdi
k_1 +k_2 +k_3 =0
\edi
I.e. $T_{\mu\nu\lambda}$ is given by the triangle diagram of Fig. 1
(the momentum routing in Fig. 1 is chosen in such a way as to ensure Bose
symmetry, see e.g. \cite{GJ1})

\medskip

$$\psannotate{\psboxscaled{450}{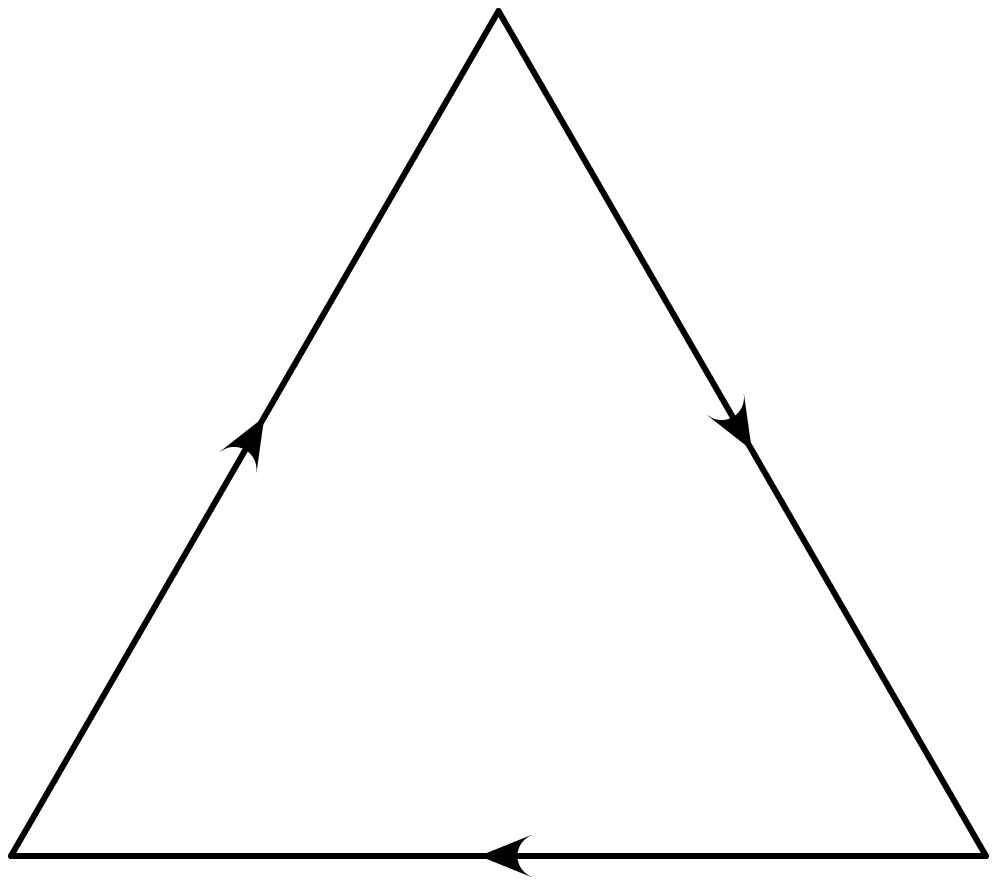}}{\fillinggrid
\at(6.2\pscm;-2.8\pscm){Fig. 1} \at(4\pscm;-1\pscm){$q-\frac{1}{3}k_1 
+\frac{1}{3}k_2$} \at(-2.3\pscm;4.5\pscm){$q-\frac{1}{3}k_3 +\frac{1}{3}k_1$}
\at(9.5\pscm;4.5\pscm){$q-\frac{1}{3}k_2 + \frac{1}{3}k_3$}
\at(-0.3\pscm;-1.1\pscm){$k_1 ,\mu$} \at(10.2\pscm;-1.1\pscm){$k_2 ,\nu$}
\at(4.1\pscm;9.1\pscm){$k_3 ,\lambda$}
}$$

\medskip

\medskip

\medskip

\medskip

\medskip

An explicit expression for $T_{\mu\nu\lambda}$ (which may be found e.g. in
\cite{And}) is
\bdi
T_{\mu\nu\lambda}(k_1 ,k_2 ,k_3)= -\frac{1}{3\pi^2}\epsilon_{\alpha\mu\nu
\lambda} \Bigl[ (k_1^\alpha -k_2^\alpha )k^2_3 I_{1,2} +(k_2^\alpha -
k_3^\alpha )k_1^2 I_{2,3} +(k_3^\alpha -k_1^\alpha )k_2^2 I_{3,1}\Bigr] -
\edi
\beq
\frac{1}{\pi^2}\Bigl[ \epsilon_{\alpha\beta\mu\nu}k_1^\alpha k_2^\beta 
k_{3\lambda} I_{1,2} +\epsilon_{\alpha\beta\nu\lambda}k_2^\alpha
k_3^\beta k_{1\mu}I_{2,3} +\epsilon_{\alpha\beta\lambda\mu}
k_3^\alpha k_1^\beta k_{2\nu}I_{3,1}\Bigr]
\eeq
\beq
I_{i,j}:=\int_0^1 dx_1 dx_2 dx_3 \frac{x_i x_j \delta (1-x_1 -x_2 -x_3)}{
k_1^2 x_2 x_3 +k_2^2 x_3 x_1 +k_3^2 x_1 x_2}
\eeq
It may be easily shown to fulfil the anomalous Ward identity
\beq
k_3^\lambda T_{\mu\nu\lambda}(k_1 ,k_2 ,k_3)=-\frac{1}{6\pi^2}
\epsilon_{\mu\nu\alpha\beta}k_1^\alpha k_2^\beta
\eeq
and analogous relations for $k_1^\mu$, $k_2^\nu$ due to Bose symmetry.

Now let us have a look at the counterterm. It is of order $e^3$ and contains 
three gauge fields. Therefore it will occur within a Feynman diagram
precisely in the same positions where the triangle graph occurs. Because
of this, we may take the counterterm $C_{\mu\nu\lambda} (k_1 ,k_2 ,k_3)$ 
into account by defining a new, ``gauge invariantly regularized''
triangle amplitude
\beq
T^g_{\mu\nu\lambda}(k_1 ,k_2 ,k_3)=T_{\mu\nu\lambda}(k_1 ,k_2 ,k_3)+
C_{\mu\nu\lambda}(k_1 ,k_2 ,k_3)
\eeq
An explicit expression for the counterterm (without charges at the three
vertices) reads
\beq
C_{\mu\nu\lambda}(k_1 ,k_2 ,k_3)=\frac{1}{6\pi^2}\Bigl[ 
\epsilon_{\alpha\beta\mu\nu}k_1^\alpha k_2^\beta \frac{k_{3\lambda}}{k_3^2}+
\epsilon_{\alpha\beta\nu\lambda}k_2^\alpha k_3^\beta \frac{k_{1\mu}}{k_1^2}+
\epsilon_{\alpha\beta\lambda\mu}k_3^\alpha k_1^\beta \frac{k_{2\nu}}{k_2^2}
\Bigr]
\eeq
Here the poles $\frac{1}{k_i^2}$ that occur in $C_{\mu\nu\lambda}$ are
described by the usual $i\epsilon$ description, i.e. they are Feynman
propagators of a massless, scalar ``ghost'' field.
It is easy to see that the new triangle amplitude $T^g_{\mu\nu\lambda}$
fulfils the naive Ward identity
\beq
k_1^\mu T^g_{\mu\nu\lambda}(k_1 ,k_2 ,k_3)=k_2^\nu T^g_{\mu\nu\lambda}
(k_1 ,k_2 ,k_3)=k_3^\lambda T^g_{\mu\nu\lambda}(k_1 ,k_2 ,k_3)=0
\eeq
Now it has to be checked whether the poles that are present in the
counterterm give actually contributions to e.g. scattering amplitudes
and thereby spoil unitarity of the theory (6).
First observe that each term in $C_{\mu\nu\lambda}$ is transverse at two
vertices and longitudinal at the third vertex,
\beq
C_{\mu\nu\lambda}(k_1 ,k_2 ,k_3)=A_{\mu\nu}(k_1 ,k_2)k_{3\lambda}+
A_{\nu\lambda}(k_2 ,k_3)k_{1\mu}+A_{\lambda\mu}(k_3 ,k_1)k_{2\nu}
\eeq
\beq
k_1^\mu A_{\mu\nu}(k_1 ,k_2)=0 \qquad \mbox{etc.}
\eeq
For a further discussion we need the gauge field propagator. The theory (6)
is gauge invariant, therefore we have to introduce a gauge fixing term as
usual and get the propagator
\beq
D^\xi_{\mu\nu}(k)=\frac{1}{k^2}(g_{\mu\nu}-\xi\frac{k_\mu k_\nu}{k^2})
\eeq
where $\xi$ is the gauge fixing parameter. We will keep $\xi$ 
arbitrary for most of our discussion, and only shortly discuss that we
can rederive our conclusions for a special choice of $\xi$.

So let us investigate Feynman graphs where triangle diagrams occur as
subdiagrams and try to find what happens with the counterterm.
First, whenever a vertex (e.g. $\lambda$) of $T^g_{\mu\nu\lambda}$ is
connected to an external photon, the part of the counterterm that is 
longitudinal at $\lambda$ vanishes, because of the transversality of the 
external photon,
\beq
\epsilon^\lambda (k_3) A_{\mu\nu} (k_1 ,k_2) k_{3\lambda}=0
\eeq
When a gauge field propagator $D^\xi_{\mu\nu}$ is connected to the
regularized triangle $T^g_{\mu\nu\lambda}$, the $\xi$-dependent part
of $D^\xi_{\mu\nu}$ does not contribute because of the Ward identity (13),
\bdi
D^\xi_{\mu' \mu} (k_1)T^g_{\mu\nu\lambda}(k_1 ,k_2 ,k_3)=
D^0_{\mu' \mu}(k_1) T^g_{\mu\nu\lambda}(k_1 ,k_2 ,k_3)
\edi
\beq
D^0_{\mu' \mu}(k)=\frac{g_{\mu'\mu}}{k^2}
\eeq
Now the gauge field propagator may connect a vertex of $T^g_{\mu\nu\lambda}$
to different types of subgraphs:
\begin{enumerate}
\item When e.g. $D^0_{\lambda \lambda'}(k_3)$ connects $T^g_{\mu\nu\lambda}$
to an external fermion-antifermion pair, the corresponding part
$A_{\mu\nu}(k_1 ,k_2)k_{3\lambda}$ of the counterterm does not contribute
because of the equations of motion for the on-shell spinors ($k_3 =p_1 +p_2$)
\beq
A_{\mu\nu}(k_1 ,k_2)k_{3\lambda}\frac{g^{\lambda\lambda'}}{k^2_3}\bar u(p_1)
\gamma_{\lambda'}\gaf v(p_2)=A_{\mu\nu}(k_1 ,k_2)\frac{1}{k_3^2}
\bar u(p_1)(\not \! p_1 +\not \! p_2)\gaf v(p_2)=0
\eeq
\item When e.g. the vertex $\lambda$ of $T^g_{\mu\nu\lambda}$ is connected 
to a closed fermion loop that is {\em not} a triangle, the counterterm
$A_{\mu\nu}(k_1 ,k_2)k_{3\lambda}$ does not contribute because of the Ward
identity for the closed fermion loop,
\beq
A_{\mu\nu}(k_1 ,k_2)k_{3\lambda}\frac{g^{\lambda\lambda'}}{k_3^2}
\wt{\langle J^5_{\lambda'}J^5_{\rho_1}\ldots J^5_{\rho_n}\rangle}
(k_3 ,p_1 ,\ldots ,p_n)=0
\eeq
\item When e.g. the vertex $\lambda$ of $T^g_{\mu\nu\lambda}$ is connected 
to a fermion line that starts and ends at an external, on-shell fermion, 
the contribution of the counterterm to an individual diagram does not
vanish. However, there are several diagrams of this type (see Fig. 2
for an example, where we denote the counterterm by a double line)

$$\psannotate{\psboxscaled{400}{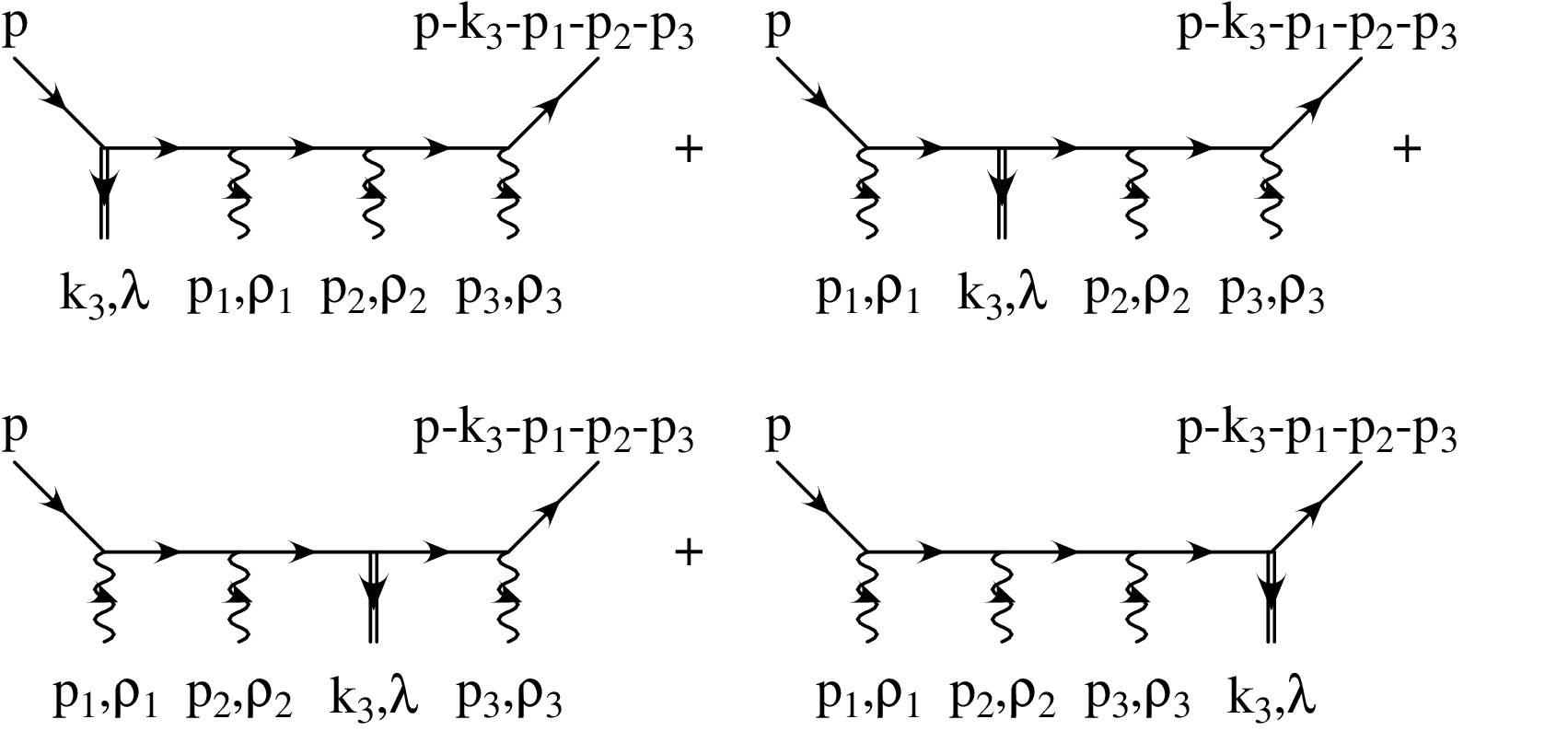}}{\fillinggrid
\at(11.5\pscm;-0.5\pscm){Fig. 2}}$$

Here it is understood that, apart from the different positions where
$k_{3\lambda}$ enters, all four diagrams (e.g. the connections of 
$\rho_1 ,\rho_2 ,\rho_3$ to the rest of the diagram) are completely
identical. The contribution of $A_{\mu\nu}(k_1 ,k_2)k_{3\lambda}$ in
this sum cancels, as may be computed easily from the expression for
Fig. 2,
\bdi
\bar u(p-p_1 -p_2 -p_3 -k_3)\cdot
\edi
\beq
\cdot \Bigl[ \gamma_{\rho_3}\gaf\frac{1}{\not \! p
\, -\not \! p_1 -\not \! p_2 -\not \! k_3}\gamma_{\rho_2}\gaf
\frac{1}{\not \! p \, -\not \! p_1 -\not \! k_3}\gamma_{\rho_1}\gaf
\frac{1}{\not \! p \, -\not \! k_3}\not \! k_3 \gaf \,  +\, \ldots \, \Bigr] 
u(p)
\eeq
by frequently using the identity
\beq
\frac{1}{\not \! p \, -\not \! k}\not \! k\frac{1}{\not \! p}=
\frac{1}{\not \! p \, -\not \! k}-\frac{1}{\not \! p}
\eeq
and the equations of motion for $u,\bar u$.

This cancellation holds for an arbitrary string of fermion propagators
beginning and ending at an external fermion. Further, this cancellation
does not depend on the remainder of the diagram that 
is connected to the other vertices $\rho_i$ of
the fermion line, therefore this cancellation continues to hold when some
other triangle diagrams or some other vertices of the same triangle
diagram (including the counterterm) are connected to the given fermion line.
\end{enumerate}

Therefore we found so far that the contribution of the conterterm 
$C_{\mu\nu\lambda}$ cancels completely as long as each vertex
$\mu ,\nu ,\lambda$ of the triangle diagram is connected either to an
external photon, or to a closed fermion loop that is {\em not} a
triangle, or to an ``open'' fermion line that begins and ends at an external 
fermion.

The last case we have to investigate is a triangle $T^g_{\mu\nu\lambda}$
that is connected to another triangle, e.g.
\beq
 T^g_{\mu\nu\lambda}(k_1 ,k_2 ,k_3)\frac{g^{\lambda\lambda'}}{k_3^2}
T^g_{\mu' \nu' \lambda'}(k_1' ,k_2' ,-k_3)
\eeq
We assume that the vertices $\mu ,\nu ,\mu' ,\nu'$ are not connected 
to triangles, therefore the corresponding parts of the counterterms vanish,
and we get ($k_3 =-k_1 -k_2 =k_1' +k_2'$)
\beq
\frac{1}{k_3^2}\Bigl( T_{\mu\nu\lambda}(k_1 ,k_2 ,k_3)+A_{\mu\nu}(k_1 ,k_2)
k_{3\lambda}\Bigr) \Bigl( T_{\mu' \nu'}{}^\lambda (k_1' ,k_2' ,-k_3)-
A_{\mu' \nu'}(k_1' ,k_2') k_3^\lambda \Bigr)
\eeq
Because $T^g_{\mu\nu\lambda}$ obeys the Ward identity (13) we may again cancel
one of the two counterterms and find e.g.
\bdi
\frac{1}{k_3^2}\Bigl( T_{\mu\nu\lambda}(k_1 ,k_2 ,k_3)+A_{\mu\nu}(k_1 ,k_2)
k_{3\lambda}\Bigr) T_{\mu' \nu'}{}^\lambda (k_1' ,k_2' ,-k_3)=
\edi
\beq
\frac{1}{k_3^2} T_{\mu\nu\lambda}(k_1 ,k_2 ,k_3)
T_{\mu' \nu'}{}^\lambda (k_1' ,k_2' ,-k_3)
+\Bigl( \frac{1}{6\pi^2 k_3^2}\Bigr)^2 \epsilon_{\alpha\beta\mu\nu}
k_1^\alpha k_2^\beta \epsilon_{\alpha' \beta' \mu' \nu'}k_1'{}^{\alpha'}
k_2'{}^{\beta'}
\eeq
and the counterterm $A_{\mu\nu}k_{3\lambda}$ does not vanish.
This term contributes e.g. to fermion-antifermion scattering via Fig. 3

$$\psannotate{\psboxscaled{500}{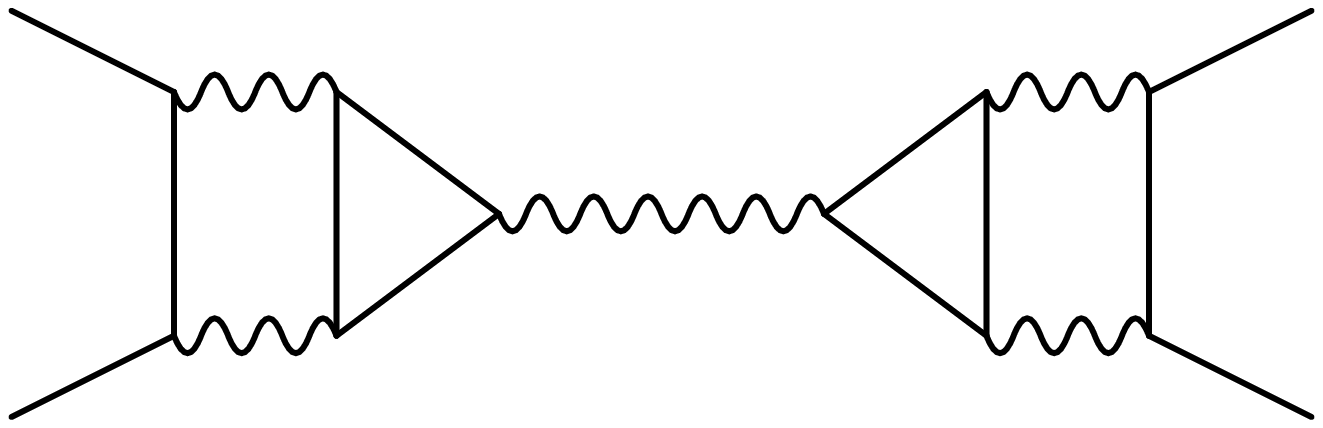}}{\fillinggrid
\at(5.5\pscm;-0.5\pscm){Fig. 3}}$$

and the counterterm certainly contributes to the imaginary part of this
scattering amplitude (remember that the $\frac{1}{k_3^2}$ are Feynman
propagators) and, therefore, violates unitarity.

Obviously, whenever two triangle diagrams are connected to each other
within a larger
graph, they will give a contribution like in (25) and therefore, in
general, violate unitarity. 

[Remark: there is a kind of cancellation that may occur in (24) for very
special values of $k_i^2$. E.g. for $k_1^2 =k_2^2 =k_3^2$ it holds that
$T_{\mu\nu\lambda}(k_1 ,k_2 ,k_3)=-C_{\mu\nu\lambda}(k_1 ,k_2 ,k_3)$ and,
therefore, the counterterm in (24) is cancelled by the triangle amplitude 
itself. However, this kind of cancellation may only occur for very specific 
values of the $k_i^2$, where the analytical structure of $T_{\mu\nu\lambda}$
simplifies to the pole structure of $C_{\mu\nu\lambda}$. For general $k_i^2$
the analytical structure of $T_{\mu\nu\lambda}$ is much more complicated,
no poles like in $C_{\mu\nu\lambda}$ occur, and there is no cancellation
(more details on the analytical structure of the triangle diagram may
be found e.g. in \cite{Ho1,Ho2}).]

Next we want to show that we find precisely the same violation of
unitarity when we fix the gauge to be Landau gauge ($\xi =1$) 
from the very beginning,
\beq
D^1_{\mu\nu}(k)=\frac{1}{k^2}(g_{\mu\nu}-\frac{k_\mu k_\nu}{k^2})
\eeq
where we see from (3), (5) that the gauge condition $\partial^\mu A_\mu
=0$ eliminates the counterterm.
 
First, when an external photon is connected to a vertex $\lambda$ of 
$T^g_{\mu\nu\lambda}$, the same cancellation of $A_{\mu\nu}k_{3\lambda}$
as above occurs.

Secondly, when all three vertices of $T^g_{\mu\nu\lambda}$ are connected to
gauge field propagators (26), the counterterm $C_{\mu\nu\lambda}$ cancels
completely because of the transversality of $D^1_{\mu\nu}$, as we already 
noticed. 
Therefore, a violation of unitarity must be due to the second term of
$D^1_{\mu\nu}$, $-\frac{k_\mu k_\nu}{k^4}$. 

Again, as long as the triangle is connected to a subdiagram that is {\em
not} a triangle, the term $-\frac{k_\mu k_\nu}{k^4}$ does not contribute,
because the second momentum (e.g. $k_\nu$) meets a transverse vertex.

On the other hand, when $D^{1\lambda\lambda'}(k_3)$ connects two triangles,
we find the contribution
\bdi
T_{\mu\nu\lambda}(k_1 ,k_2 ,k_3)\frac{-k_3^\lambda k_3^{\lambda'}}{k_3^4}
T_{\mu' \nu' \lambda'}(k_1' ,k_2' ,-k_3)=
\edi
\beq
\Bigl(\frac{1}{6\pi^2 k_3^2}\Bigr)^2 
\epsilon_{\alpha\beta\mu\nu}k_1^\alpha k_2^\beta
\epsilon_{\alpha' \beta' \mu' \nu'}k_1'{}^{\alpha'}k_2'{}^{\beta'}
\eeq
and, therefore, the same unitarity violating term as before.

\section{Alternative formulation}

Finally, we want to discuss an alternative, equivalent formulation of the 
theory (6) that was given in \cite{Fed}, too, 
and show how the violation of unitarity
occurs there. The alternative formulation of the theory is given by the
Lagrangian
\bdi
L=\bar \Psi (i\dsla +e\Asla \gaf )\Psi -\frac{1}{4}F_{\mu\nu}F^{\mu\nu}
+\frac{e^3}{96\pi^2}\epsilon_{\mu\nu\alpha\beta}F^{\mu\nu}F^{\alpha\beta}
(a+b)-
\edi
\beq
\frac{\mu^2}{2}(\partial_\mu a -A_\mu )^2 +\frac{\mu^2}{2}(\partial_\mu b
-A_\mu )^2
\eeq
Here $a$ and $b$ are scalar fields that only occur as internal lines in Feynman
diagrams ($\mu$ is a dimensionful parameter on which the theory does not 
depend).

This theory is gauge invariant provided that, in addition to the usual gauge 
transformations
\beq
A_\mu \ra A_\mu +\partial_\mu \lambda \quad ,\quad \Psi \ra e^{ie\lambda\gaf}
\Psi \quad ,\quad \bar \Psi \ra \bar \Psi e^{ie\lambda \gaf}
\eeq
the fields $a$, $b$ transform according to
\beq
a\ra a+\lambda \quad ,\quad b\ra b+\lambda
\eeq
The original, nonlocal action (6) may be recovered from (28) by performing
the (Gaussian) path integration over $a$ and $b$.

In \cite{Fed} it was claimed that the ``ghosts'' $a$, $b$ will give no net
contribution to physical amplitudes. There is indeed a partial cancellation
of the $a$ and $b$ field contributions, 
because their kinetic terms have opposite signs and lead therefore to
propagators with opposite signs
\beq
D_{{a\atop b}}(k)= \frac{\pm 1}{\mu^2 k^2}
\eeq
The $a$, $b$ in (28) have two interaction terms. They couple to the index 
density $\epsilon_{\mu\nu\alpha\beta}F^{\mu\nu}F^{\alpha\beta}$ with equal
couplings $\frac{e^3}{96\pi^2}$, and to the gauge field (more precisely, to
$\partial^\mu A_\mu$) with opposite couplings $\mp \mu^2$. Therefore, a
cancellation between $a$ and $b$ propagators occcurs as long as the $a$
($b$) propagators connect either two index densities or two gauge fields.
On the other hand, when the $a$ ($b$) propagators connect one index density 
and one gauge field, the opposite signs of the propagators are compensated
by the opposite signs of the coupling to the gauge field, and the two
contributions have equal sign. When these two terms are taken into account
within a Feynman diagram, they lead precisely to the unitarity violating terms
that we found in the first formulation of the theory,
\bdi
2\frac{1}{96\pi^2}\epsilon_{\mu\nu\alpha\beta}F^{\mu\nu}(x)F^{\alpha\beta}(x)
a(x) \mu^2 \partial^\lambda_y a(y) A_\lambda (y) \longrightarrow
\edi
\beq
\frac{1}{6\pi^2}\Bigl( \epsilon_{\alpha\beta\mu\nu}k_1^\alpha k_2^\beta
\frac{k_{3\lambda}}{k_3^2} +\epsilon_{\alpha\beta\nu\lambda}k_2^\alpha
k_3^\beta \frac{k_{1\mu}}{k_1^2} +\epsilon_{\alpha\beta\lambda\mu}
k_3^\alpha k_1^\beta \frac{k_{2\nu}}{k_2^2}\Bigr)
\eeq
where we inserted the propagator (31) for the contraction of the two $a$
fields and took into account all the possibilities to contract
the three gauge fields in (32) with three fixed gauge fields of the
remainder of the diagram where (32) is inserted. We recover precisely the
counterterm (12).

\section*{Acknowledgement}

The author thanks R. Jackiw for suggesting the problem and for very
helpful discussions.

This work is supported by a Schr\"odinger stipendium of the Austrian FWF.

\medskip

\medskip

\medskip

{\bf Note added:} In a recent answer \cite{Fed2} 
to our paper it has been claimed that
the counterterm in (25) is cancelled by a contribution of the first term
(the triangle part) and, therefore, a violation of unitarity does not occur.
Although this point has already been discussed in the literature \cite{And},
we nevertheless want to comment on it briefly. For a clear distinction between
gauge field propagator and counterterm (and to be close to the notation in
\cite{Fed2}) we introduce a small gauge field mass $m$ and rewrite our eq. (25)
\bdi
\frac{1}{k_3^2 -m^2}T_{\mu\nu\lambda}(k_1 ,k_2 ,k_3)T_{\mu' \nu'}^{\lambda '}
(k_1' ,k_2' -k_3) + \frac{1}{(6\pi^2)^2}\frac{1}{(k_3^2 -m^2)k_3^2}
\epsilon_{\alpha\beta\mu\nu}k_1^\alpha k_2^\beta
\epsilon_{\alpha'\beta'\mu'\nu'}k_1'{}^{\alpha'}k_2'{}^{\beta'} .
\edi
Now the claim in \cite{Fed2} is that the first term, too, contains a pole
$\frac{1}{k_3^2}$ that precisely cancels the second term (the counterterm).
This is indeed the case at the symmetric point $k_1^2 =k_2^2 =k_3^2$, where
each triangle is exactly equal to minus the counterterm, $T_{\mu\nu\lambda}
=-C_{\mu\nu\lambda}$. Therefore, at $k_1^2 =k_2^2 =k_3^2 =0$ the imaginary
part of the counterterm ($\sim \delta (k_3^2)$) is cancelled in the unitarity 
relations.

However, this cancellation does not hold in the general case. For $k_1^2 ,
k_2^2 \ne k_3^2$ the triangle graph does not contain a $\frac{1}{k_3^2}$ pole
and, therefore, its imaginary part does not contain a $\delta (k_3^2)$ term,
as may be shown by a straight forward but tedious calculation. For the
analogous case of the VVA triangle graph the computation of the imaginary part
of $T_{\mu\nu\lambda}^{\rm VVA}$ has been performed e.g. in \cite{Ho2,Frish}.
The explicit expression for the imaginary part of  
$T_{\mu\nu\lambda}^{\rm VVA}$ (which is a rather complicated function) does
not contain a $\delta (k_3^2)$ term in the general case; only in the limit
$k_1^2 ,k_2^2 \to 0$ such a $\delta(k_3^2)$ term is produced, see
\cite{Ho2,Frish}. 

[The physical reason for this behaviour of the imaginary
part, which stems from the cutting of the triangle graph, 
may be easily understood: when all momenta squared entering the triangle are
equal, the fermions of the triangle have to be collinear after the cutting,
giving thereby rise to a $\delta (k_3^2)$ imaginary part; on the other hand,
once not all of the $k_i^2$ are equal, the fermions need not be collinear
after the cutting and, therefore, produce the usual discontinuity above
the real particle production threshold of the complex $k_3^2$ variable, see 
\cite{Frish,CG1}.]  

Therefore, there is {\em no} cancellation of the counterterm for $k_1^2 ,
k_2^2 \ne k_3^2$. Further, there are, of course, contributions to the unitarity
relations of some scattering processes where $k_1^2 ,k_2^2 \ne 0$ (e.g.
for our Fig. 3). In all such cases the $\delta (k_3^2)$ term of the 
counterterm cannot be cancelled by a contribution from the triangle graph,
and unitarity is violated.

\end{document}